\documentclass{elsart}
\usepackage{graphicx}
\usepackage{harvard}
\begin{document}
\raggedleft{IPJ 2004/01}
\begin{frontmatter}
\title{{\em Pi of the Sky}\thanksref{pi_url} --- all-sky, real-time search for fast optical
    transients\thanksref{KBN}}
  \thanks[pi_url]{http://grb.fuw.edu.pl}
  \thanks[KBN]{This work is supported by the Polish Committee for Scientific Research
   under grant 2 P03B 038 25}
\author[PW]{A. Burd}, \author[IFD]{M. Cwiok}, \author[IFD]{H. Czyrkowski},
\author[IFD]{R. Dabrowski}, \author[IFD]{W. Dominik}, \author[PW]{M. Grajda},
%\author[IPJ]{M. Gorski},
\author[PW]{M. Husejko}, \author[PW]{M. Jegier},
\author[PW]{A. Kalicki}, \author[PW]{G. Kasprowicz}, \author[IFD]{K. Kierzkowski},
\author[UKSW]{K. Kwiecinska}, \author[CFT]{L. Mankiewicz}, \author[IPJ]{K. Nawrocki},
\author[OA]{B. Pilecki}, \author[IFD]{L.W. Piotrowski}, \author[PW]{K. Pozniak},
\author[PW]{R. Romaniuk}, \author[PW]{R. Salanski}, \author[IPJ]{M. Sokolowski},
\author[OA]{D. Szczygiel}, \author[IPJ]{G. Wrochna\corauthref{cor1}},
\corauth[cor1]{Corresponding author.} \ead{wrochna@fuw.edu.pl}
and \author[PW]{W. Zabolotny}
\address[PW]{Institute of Electronic Systems, Warsaw University of Technology,
             Nowowiejska~15/19, 00-665~Warsaw, Poland}
\address[IFD]{Institute of Experimental Physics, Faculty of Physics, Warsaw University,
              Ho\.za~69, 00-681~Warsaw, Poland}
\address[UKSW]{Department of Mathematics and Natural Sciences, Cardinal Wyszynski University,
               Dewajtis~5, 01-815~Warsaw, Poland}
\address[CFT]{Center for Theoretical Physics, Polish Academy of Science, 
              Al.~Lotnik\'ow~32/46, 00-668~Warsaw, Poland}
\address[OA]{Astronomical Observatory, Warsaw University, Al.~Ujazdowskie~4,
             00-478~Warsaw, Poland}
\address[IPJ]{Soltan Institute for Nuclear Studies, Ho\.za~69, 00-681~Warsaw, Poland}

\begin{abstract}
An apparatus to search for optical flashes in the sky is described.
It has been optimized for gamma ray bursts (GRB) optical counterparts.
It consists of $2\times 16$ cameras covering all the sky.
The sky is monitored continuously and the data are analysed on-line.
It has self-triggering capability and can react to external triggers with negative delay.
The prototype with two cameras has been installed at Las Campanas (Chile)
and is operational from July 2004.
The paper presents general idea and describes the apparatus in detail.
Performance of the prototype is briefly reviewed and perspectives for the future are outlined.
\end{abstract}

\end{frontmatter}

\section{Introduction}
\label{intro}
Gamma ray bursts (GRB) are one of the most intriguing phenomena in the Universe.
Several thousands of them have been observed by gamma ray detectors carried by satellites.
In order to understand their nature, observations should be carried out in all wavelengths. Optical domain is especially important, because of high spatial resolution and possibility
of detailed spectral analysis.
Unfortunately, only a few dozens of GRB sources have been seen in visible light.
Moreover, most of them have been observed as late as hours or days after the burst
as very faint afterglows.
So far, only one GRB was caught by an optical device within the first minute
after a GRB \cite{GRB990123}.
The observation was made by a small robotic telescope ROTSE equipped with photo-lenses
of 10\,cm aperture \cite{ROTSE}.
Typical professional telescopes have too large inertia and too small field of view
to be able to react promptly to GRB alerts from satellites.

Since then, many small robotic telescopes have been build and
installed around the globe in order to search for GRB optical
counterparts. Representative examples are BOOTES \cite{bootes}, MASTER \cite{master}, 
RAPTOR \cite{raptor}, and ROTSE-III \cite{rotse3}.
Recent review was written by \cite{robo_rev}. 

Typically, they have optics of rather
short focal length and cover relatively large field of view
(1-100$^{\circ^2}$). They await for GRB alerts distributed by the GCN
system \cite{GCN} and move to the target as soon as possible in
order to take exposures. During long periods between GCN alerts
these systems typically observe a chosen Field Of View (FOV) and
collect data in search for transient phenomena of astrophysical
interest.

Unfortunately, so far the GRB930123 recorded by ROTSE remains the only
optical observation at the time comparable to the duration of the
gamma burst itself. The reason is two-fold. First, decision making
process and propagation of the triggering information from a
satellite to the observing device takes time. Second, the inertia of
the devices, although much smaller than that for large telescopes,
is still responsible for a significant delay. During this time the
object in question is fading rapidly and it goes beyond the limiting
magnitude of the device before the first exposure begins. It is
expected that alert system associated with the "Swift" satellite
\cite{Swift}, scheduled for launch this year, will operate more
efficiently and limit delay related to signal propagation from
satellite detector to GCN. However, even the fastest systems cannot
guarantee systematic observations of the critical region of the sky
prior to the satellite trigger. Certainly, such observations may
reveal additional phenomena associated with GRB and shed a new
light on their physical nature \cite{BP}.

\section{The idea of the real-time search with multilevel trigger system}
\label{idea}
In order to overcome the two major problems of classical robotic telescopes we propose
a different approach, based on author's experience from particle physics experiments.
Trigger propagation time can be eliminated if the device has self-triggering capability.
Inertia of the system does not matter if the object in question is already inside FOV
before the investigated phenomenon (e.g. GRB) takes place.
The apparatus described in this paper exhibits those two features.

We propose to build a system consisting of a number of CCD cameras covering as wide field
of view as possible.
The cameras monitor continuously the sky taking relatively short (5-10\,s) exposures.
The data are analyzed on-line, in search for optical transients.
The idea is simple: it is enough to check for a presence of a star-like object
in a given frame, which was not present in preceding frames.
However, practical realization is difficult, because of large data stream involved
(see an example below).
It is impossible to invent a single algorithm, which is fast enough,
and has high efficiency and low rate of false triggers at the same time.
The problem could be solved by implementing a multi-level trigger system.
The first level algorithms are very simple and have high efficiency for interesting events,
but they produce a lot of background.
The rate of background events at this stage can be several orders of magnitude higher
than the rate of real events we are looking for.
The only purpose of the first level algorithms is to reduce the data stream to be analyzed
by higher levels.
Thus, the second level algorithm can be somewhat more complicated and perform better
background rejection.
The highest levels deal only with a very low rate of suspected events and can employ
sophisticated algorithms to clean up the final sample.

Selected events can be submitted to larger telescopes to follow and/or can be checked
against GRB triggers from other sources, even if they come much later.
The data cannot be stored for a long time, because of limited disc space,
but can be temporarily kept in some buffer, to examine any late arriving external alerts.

\section{"$\pi$ of the Sky" apparatus design}
\label{design}
Practical realization of the ideas outlined in the previous section may look as follows.
One observing module consists of a $4\times 4$ matrix of CCD cameras.
The camera is based on CCD with $2000\times 2000$ pixels $15\times 15\,\mu$m$^2$ each.
Equipped with lenses of focal length $f=50$\,mm, the camera covers $33^\circ\times 33^\circ$
field of view. The module of 16 cameras covers almost all visible sky down to $20^\circ$
above the horizon.
This is a solid angle above $\pi$ steradians and hence the name of the apparatus ---
"{\em$\pi$ of the Sky}".

The most common background sources for flash recognition algorithms
are cosmic rays crossing the CCD and sunlight reflexes from
artificial satellites. In order to fight this background two twin
modules should be installed at some distance of the order of 10\,km.
Cosmic rays can be eliminated by taking simple coincidence. Flashing
satellites could be resolved by their parallax.

Gamma ray bursts typically have duration of 0.01-100\,s.
GRB930123 flash observed by ROTSE also had the decay time of the order of dozens of seconds.
Therefore, the exposure time of 2-10\,s seems to be optimal to search for GRB related flashes.
With the focal length of  $f=50$\,mm and the pixel size of 15\,mm one can take 5\,s
exposures on a static mount.
This is very convenient and significantly reduces the cost.
The CCD sensitive area is $30\times 30$\,mm$^2$ and can be illuminated
with standard camera lenses, commercially available for a low price.

With 5\,s exposures and 1\,s readout time one can take 10 frames per minute,
i.e. 5000-10000 frames per night.
This imposes severe requirement for the mechanics of a shutter,
which for a few years of operation must sustain $10^7$ opening cycles.
Single frame from one camera occupies 8\,MB (megabytes).
The average data rate from a single camera is thus 1.3\,MB/s = 3\,GB/h.
Two modules, 16 cameras each, produce the data stream of 100\,GB/h, i.e. 1\,TB/night.
This poses the limit on the temporary data storage.
Most of the analysis must be done on-line, in real time.
Only carefully selected data can be kept for further inspection.

Realisation of the project outlined above is foreseen in the following steps:
\begin{itemize}
 \item Phase-0. Tests with one and with two cameras on a fixed mount.
 \item Phase-1. Two cameras observing the same field, installed on a robotic mount
 \item Phase-2. Two modules, 16 cameras each, as described above
 \item Phase-3. Four pairs of modules installed around the globe,
                to maximise the coverage and observing time
\end{itemize}

\section{Phase-0 tests}
\label{phase0}
\begin{figure}[hbtp] 
  \begin{center}
    \begin{tabular}{p{.28\textwidth}p{.68\textwidth}} 
      \includegraphics[width=\linewidth]{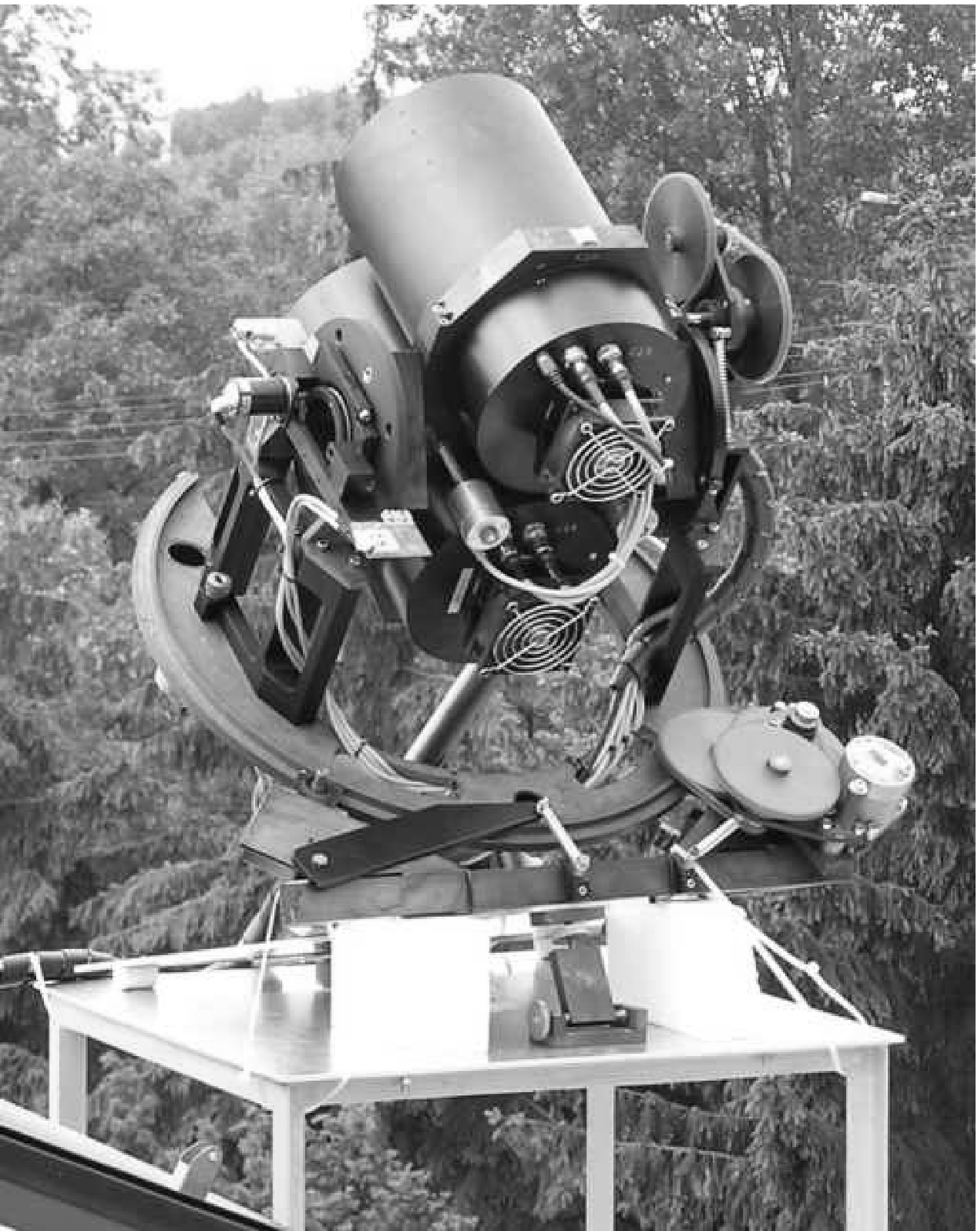} &
      \includegraphics[width=\linewidth]{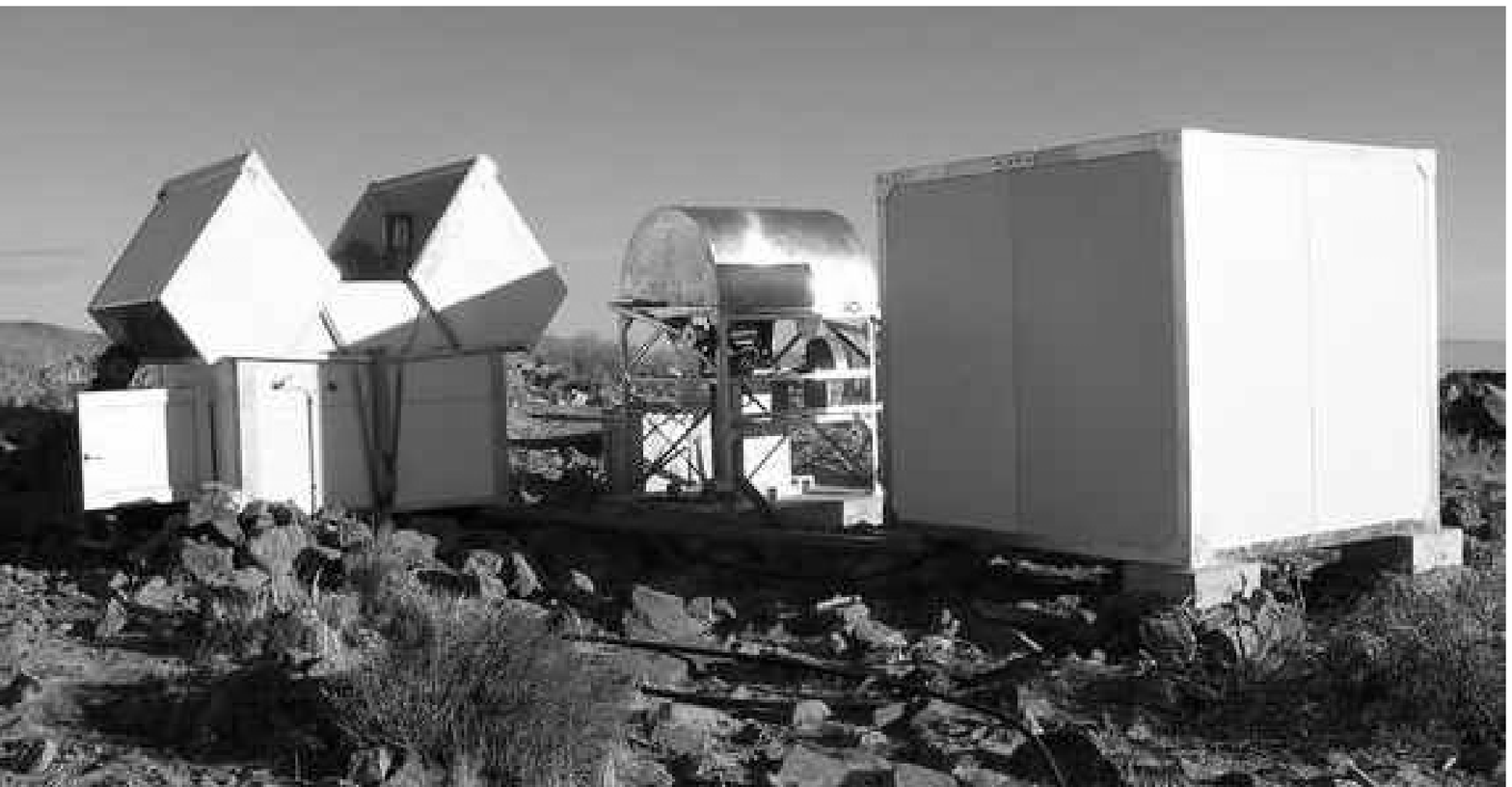} \\[-6mm]
      \caption{The "{\em $\pi$ of the Sky}" cameras and the mount at the test site in Poland.}
      \label{fig:brw} &
      \caption{The Las Campanas site. Left to right: ASAS dome housing 
         "$\pi$ of the Sky" apparatus, ASAS 10" telescope dome, Control Room.}
      \label{fig:lco} \\[-6mm]
    \end{tabular}
  \end{center} 
\end{figure}

The phase-0 of the project has been already completed.
Tests have been performed at Brwinow (52.14725$^\circ$\,N, 20.71850$^\circ$\,E),
30\,km West of Warsaw.
We have started with a modified G-16 camera \cite{G16} equipped with a Kodak CCD KAF-0401E.
It has $768\times 512$ pixels, $9\times 9\,\mu$m$^2$ each.
Images from about 50 nights between November 2002 and September 2003 have been recorded.
In the meantime the first prototype of $2000\times 2000$ pixel camera has been built.
It was used for further tests from October 2003. The first final camera for the phase-2 has been completed in February 2004, whereas the second one in June 2004 (Fig.~\ref{fig:brw}).

Almost 200 GB of data have been collected in different
configurations. Those include both, fixed and robotic 
mounts. The data have been used thoroughly to design and tune flash
recognition algorithms. Major background sources have been
identified and their rates have been measured. At least one
interesting fast optical transient has been observed. The results
will be published elsewhere.

\section{"{\em $\pi$ of the Sky}" phase-1 system}
\label{phase1}
The phase-1 system has been installed at the Las Campanas Observatory in Chile
in June 2004 (Fig.~\ref{fig:lco}).
Regular operation started in July 2004.
The system consists of two custom designed CCD cameras installed on a robotic mount (Fig. 1).

The cameras are based on CCD442A sensor by Fairchild Imaging.
The CCD has $2032\times 2032$ sensitive pixels, $15\times 15\mu$m$^2$ each.
It is read out with a frequency 2\,MHz/pixel, so the entire matrix is read out in 2\,s.
After amplification, the signal is digitized by 16-bit ADC and stored in a memory.
The camera is read out and controlled by a PC through a fast USB 2.0 interface.
Data transfer takes less than one second and can be done while the next exposure
is already being taken.
Readout speed, amplifier gain and other parameters are programmable via USB.
The sensor is cooled with a stack of two Peltier modules about 35 degrees below
the ambient temperature.
Special heavy-duty mechanical shutter was designed to sustain over $10^7$ opening cycles.
A prototype has been tested at high frequency in a lab.
The first signs of degradation have appeared after $1.2\cdot 10^7$ cycles.
Details of the camera design will be published elsewhere.

Each camera is equipped with Planar-T* photo-lenses by Carl-Zeiss of
a focal length $f=50$\,mm and an aperture $d = f / 1.4$ . Focusing
is performed by a step motor with a controller build into the
camera and remotely controlled through the USB. The effective field
of view is $33^\circ\times 33^\circ$. The two cameras are installed
on a common mount in such a way that they observe the same field.
The mount we use was originally designed for the ASAS experiment
\cite{ASAS} and modified to suit the "{\em $\pi$ of the Sky}" needs.

It is driven by step motors controlled by a PC through RS232 serial interface.
The mount can reach any point in the sky in less than one-minute time.
The whole setup is installed in the ASAS dome together with other ASAS telescopes.

\section{Operation and data flow}
\label{operation}
The apparatus is controlled by a PC located inside the dome.
Second PC, located in a nearby Control Room is used for off-line data analysis.
The system is fully autonomous, but also fully controllable via Internet.
During the normal operation the system runs autonomously according to the
preprogrammed schedule. Dedicated script language has been developed to make
the schedule programming easy and flexible.

For most of the time the cameras follow the field of view of the
HETE satellite \cite{HETE}. Its position is read out from the
Internet in regular intervals and the mount position is
automatically corrected accordingly. If the HETE FOV is not visible,
another location in the sky is programmed. The system is also
listening to GCN alerts received directly and through another server
in Warsaw as a backup. Should an alert located outside the current
FOV arrive, the mount automatically moves towards the target and
exposures are being taken. Twice a night an all sky scanning is
performed. 16 fields are visited and three images of each are taken by
both cameras. A single scan last about 20 minutes.

Because the mount follows the sideral movement, one can take longer exposures
than in the case of a fixed system.
We have chosen 10\,s as a compromise between the magnitudo reach and time resolution
for short flashes.
The images are immediately analyzed while in the computer RAM in search for flashes
with a rise time of the order of seconds.
Then, they are temporarily stored on a disc and can be reexamined in case of late
arrival of an external alert.
If a flash candidate is found the $100\times 100$ pixel samples of $\pm 7$ frames
are stored permanently for the record.

In the meantime, the images are copied to the second PC, which
superposes the images and searches for optical transients with a
rise time of minutes. During the day, two analyses are performed in
parallel on the temporarily stored data. The first PC runs fast
photometry on individual frames, which can be used later to study
rapidly varying objects. The second PC performs precise photometry
\cite{ASAS}  on images superposed by 20. This could be used to study
variable stars etc. The results are stored permanently on a disc.
Out of almost 30\,GB of data taken every night, about 2\,GB of
results is stored permanently. After 2-3 months a 200\,GB removable
disc with the results is replaced and taken to Warsaw for further
analysis.

\section{Software}
\label{software} Both PC are running the Linux operating system. The
software is mostly custom written in C++. It consists of a number of
modules taking care of different devices: the mount, the cameras,
the data acquisition system, the GCN server, etc. All modules are
governed by the central module "{\tt piman}" scheduling the tasks
and controlling the information flow. The communication between
modules it based on {\tt CORBA}. Image processing classes make use of
"{\tt cfitsio}" package \cite{fitsio}. Flash recognition algorithms
and fast photometry are custom developed, whereas precise photometry
and astrometry is adopted from ASAS \cite{ASAS}.

Monte Carlo technique was extensively used in development of the algorithms.
Optical flashes were simulated by cutting star images from real sky images
and pasting them into images under study at random locations.
This method was used to evaluate efficiency of flash detection algorithm.
Details are described in \cite{simulation}.

The flash recognition algorithm compares a given frame with several preceding ones.
It searches for a star-like object, which is missing on preceding frames.
Most of the false events are caused by cosmic rays, but those are easily eliminated
by coincidence of the two cameras.
The most severe background is due to the sunlight reflexes from artificial satellites.
We try to eliminate those in two ways.
First, we search for aligned flashes in a single frame, or in different frames.
Second, we check the time and location of the flash against a database.
Every evening a fresh database is built by merging several ones available
on the Internet \cite{mccants}.

\section{System reliability}
\label{reliability}
The apparatus operates at the Las Campanas Observatory without a permanent human supervision.
Therefore, the system must be very reliable.
It was achieved by employing self-diagnostics and remote monitoring, as well as hardware
redundancy and flexible configuration.

Special care has been taken to ensure seamless recovery after a system failure without
human intervention at the site.
The cameras have a hardware watchdog build in, which automatically resets the camera
and thus reestablishes connection with the PC in case of protocol failure.
In addition to the direct Internet connection the two PC are connected together with
a Gbit Ethernet.
It ensures fast data transfer between the two PC and also serves as a backup
in case one of the direct connections breaks.
Both PC have "{\tt Wake on LAN}" and "{\tt Boot from LAN}" capabilities and can be
started from the net in case of a file system failure.
Remote file system could be installed and used to recover the local one.
Each computer can be remotely reset or powered down/up by the relays driven by the second one.
This feature has proved itself to be extremely useful in case of system hang-ups.

The system is designed to be immune to network failures.
It can run autonomously for many days.
Each night it reads the current schedule from a script and a default script is executed
if no current script is found.
The system can be effectively monitored with a very small bandwidth.
Direct communication with the system is provided by a console module "{\tt pishell}"
which can run at any place and interact with the system exchanging short {\tt COBRA} packets.
Basic information about the system ($<\,20$\,kB) is automatically copied every 10\,min
to a WWW server in Warsaw.
Selected sky images are compressed to 4\,kB JPEG and also copied to the Warsaw
server every 20\,min.
Should any failure occur, the system sends an SMS with the appropriate information
to a mobile phone of a person in charge.

\section{System performance}
\label{performance}
The phase-1 "{\em $\pi$ of the Sky}" system operates regularly from July 2004.
Till the end of October, three nights were lost for the maintenance after
a double disk failure and several more due to weather conditions.
About half a million sky images have been taken, accounting for 4 terabytes of data.
Each frame contains about 20\,000 stars and the results consists of 10 billion
photometric measurements.

\begin{figure}[hbtp] 
  \begin{center}
    \begin{tabular}{p{.47\textwidth}p{.47\textwidth}} 
      \includegraphics[width=\linewidth]{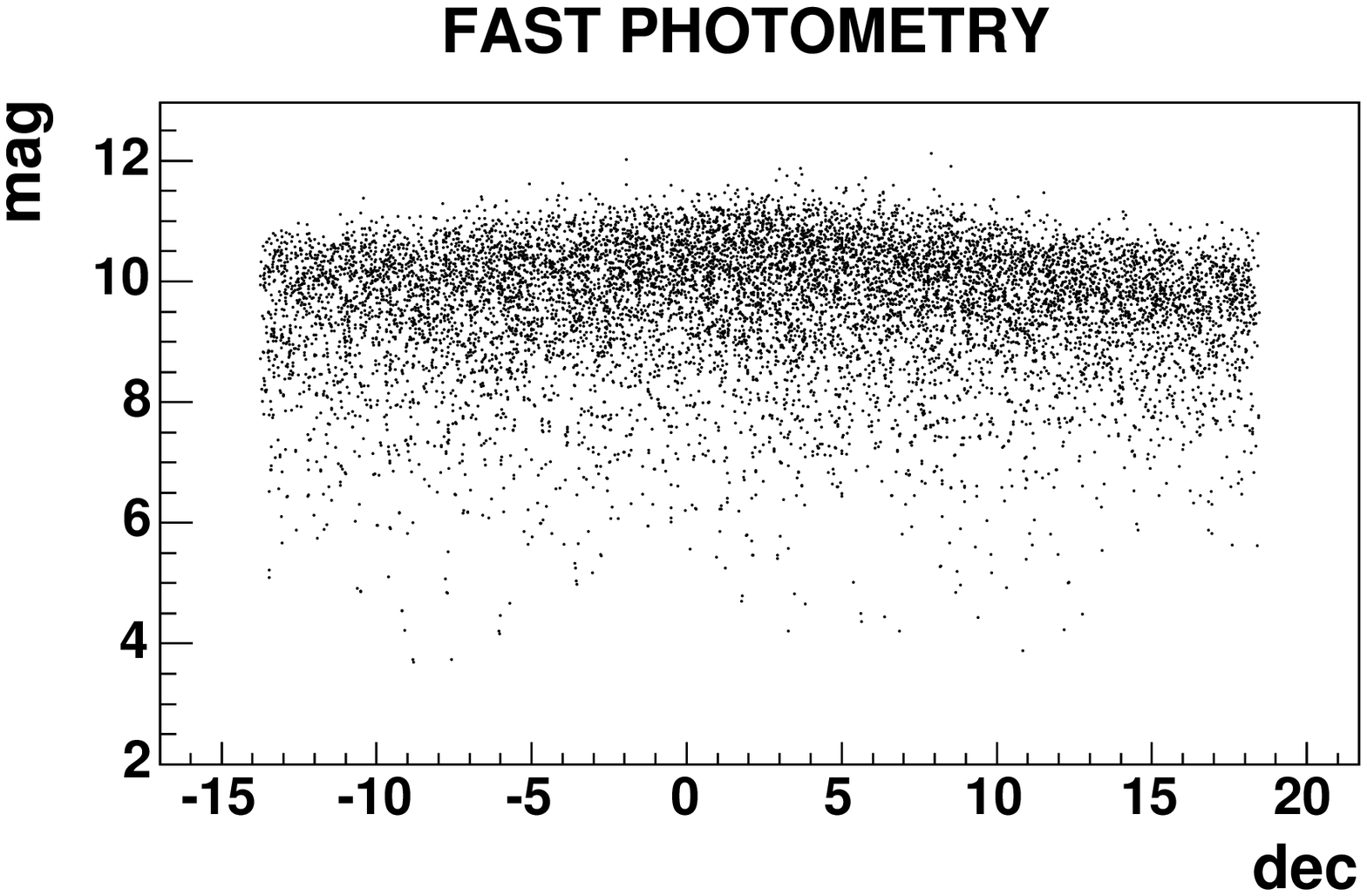} &
      \includegraphics[width=\linewidth]{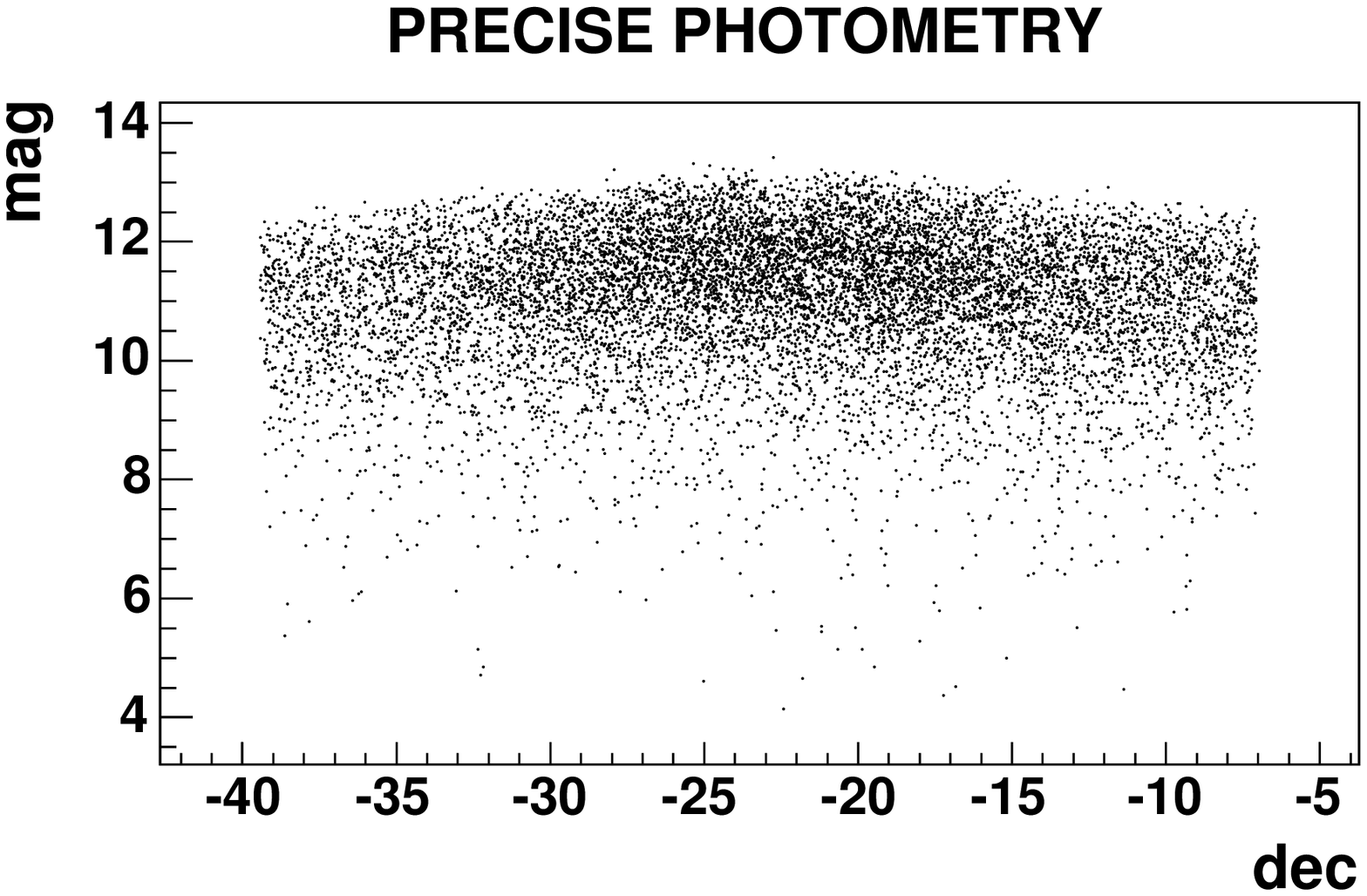} \\[-6mm]
      \caption{Fast photometry magnitudo of stars vs position in the frame (degrees).}
      \label{fig:magdecff} &
      \caption{Precise photometry magnitudo of stars vs position in the frame (degrees).}
      \label{fig:magdec} \\[-6mm]
    \end{tabular}
  \end{center} 
\end{figure}

The limiting magnitude is $10-11^m$ for single frames and $11-12^m$ for frames coadded
by 20 (Fig.~\ref{fig:magdecff} and~\ref{fig:magdec}).
Exact limit depends on the sky background, which strongly varies with the Moon phase.
It also depends on the position in the frame.
The corners are somewhat less illuminated than the center.

During the first three months of the regular operation, several short optical flashes
have been detected.
Most probably they are caused by sun reflexes from artificial satellites,
although they do not correspond to any satellite listed in commonly available databases.
One optical transient lasting $>\,10$\,s was observed during tests with a single camera,
which cannot be explained by a satellite reflex.
So far, no optical transient coinciding with a known GRB was found.
Limits for optical counterparts have been given for a few GRB \cite{Pi-GCN}.
Detailed results will be published elsewhere.

\section{Perspectives}
\label{perspectives}
\begin{figure}[hbtp] 
  \begin{center}
    \begin{tabular}{p{.47\textwidth}p{.47\textwidth}} 
      \includegraphics[width=\linewidth]{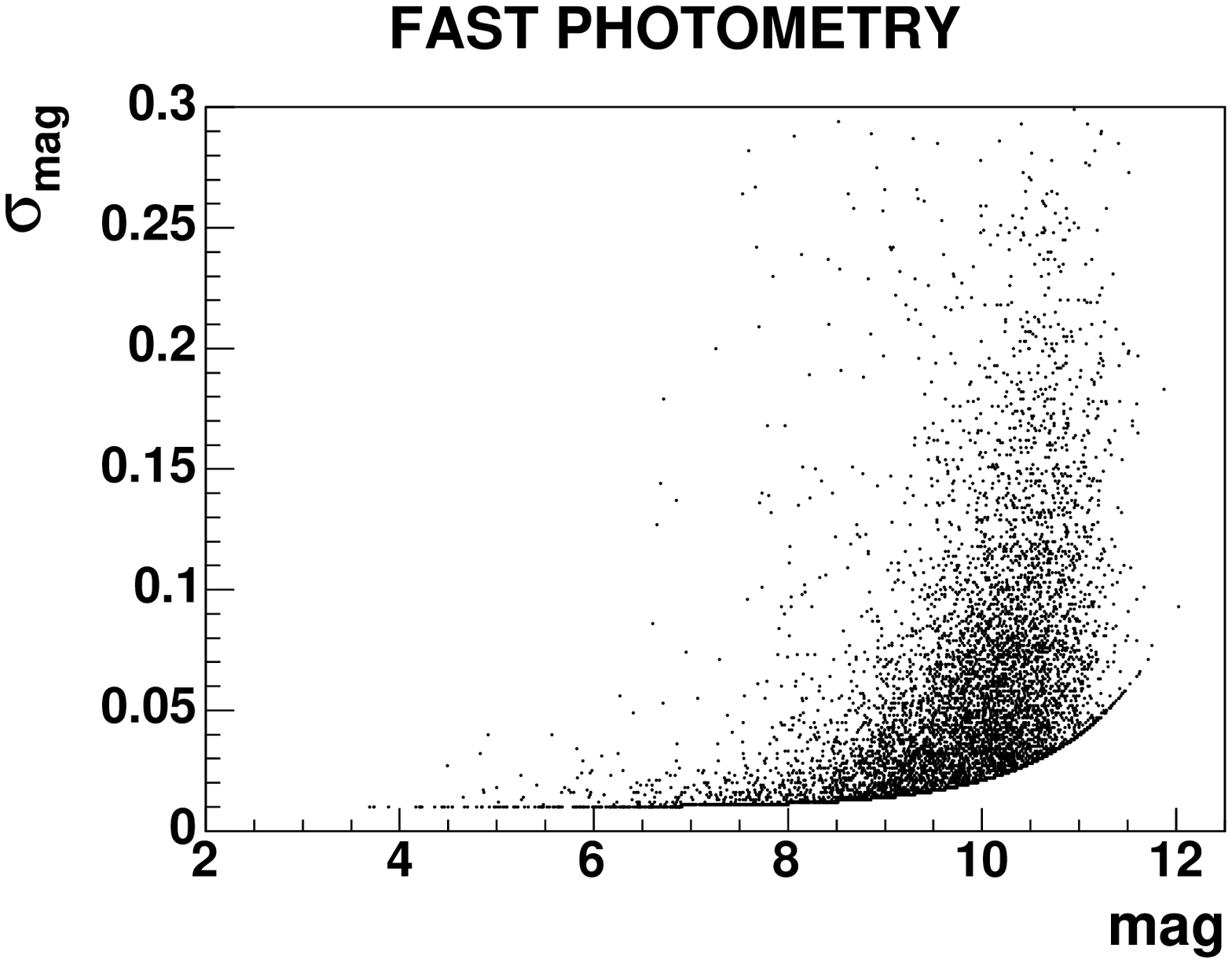} &
      \includegraphics[width=\linewidth]{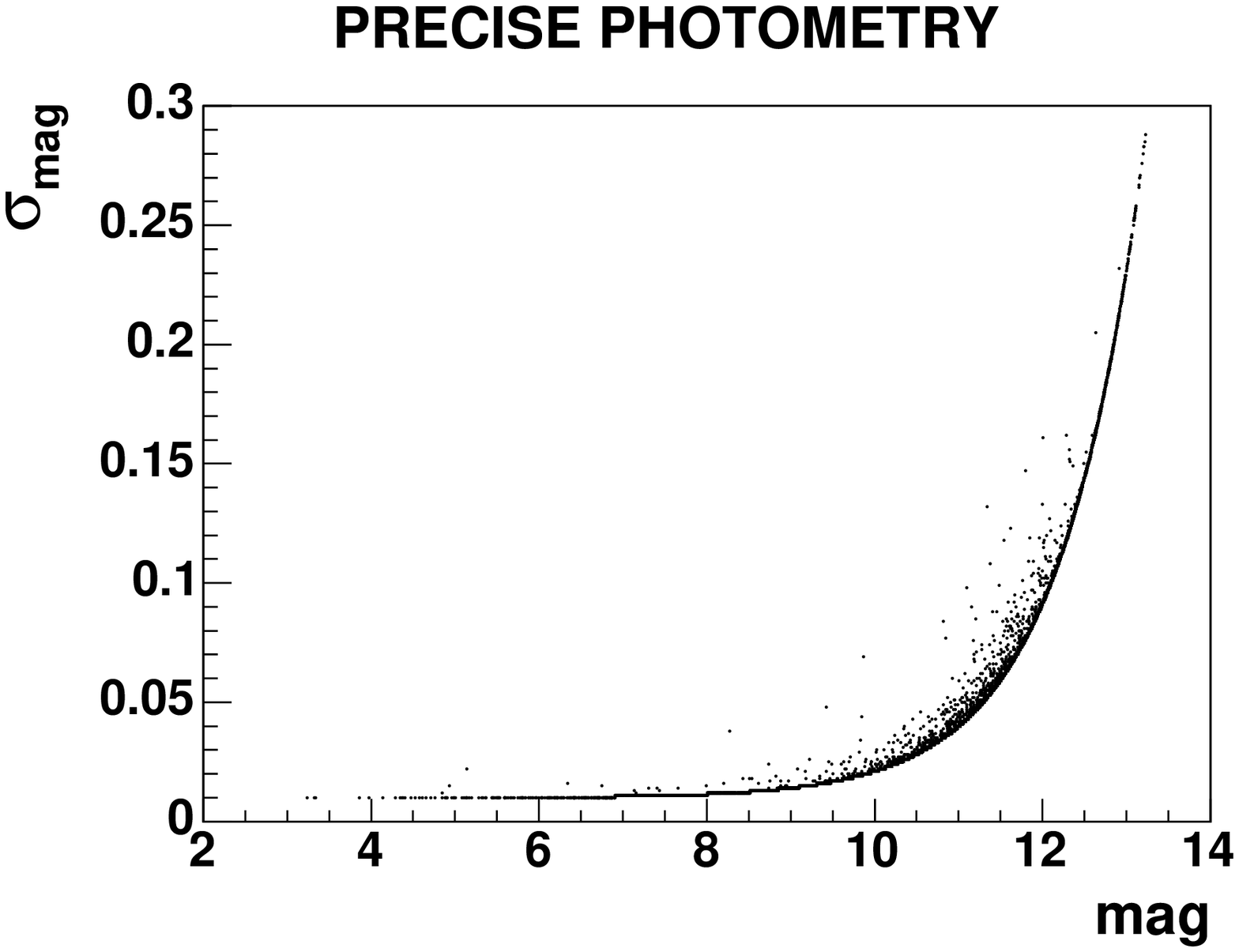} \\[-6mm]
      \caption{Precision vs brightness for the fast photometry.}
      \label{fig:sigmagff} &
      \caption{Precision vs brightness for the precise photometry.}
      \label{fig:sigmag} \\[-6mm]
    \end{tabular}
  \end{center} 
\end{figure}

Current efforts in the project concentrate in two areas.
First, the software is still being developed to improve flash recognition algorithms
and increase precision of the photometry (Fig.~\ref{fig:sigmagff} and~\ref{fig:sigmag}).
The work is going on to develop a database with photometric measurements,
which can be used to study variable stars, etc.
Second, the design of the phase-2 apparatus ($2\times 16$ cameras covering all the sky)
is moving from a conceptual to a technical level.
At the same time we search for optimal locations of the phase-3 systems.

The "{\em $\pi$ of the Sky}" apparatus can be also considered as an Earth based prototype
of a lighter, yet more powerful system, which can operate in the space \cite{astral}.
On the other hand it stands as a milestone towards future telescope "farms" monitoring
all sky in search for rare and rapid phenomena.
The system is already integrated with GCN \cite{GCN} and OTC \cite{OTC} networks.
"{\em $\pi$ of the Sky}" triggers will be soon automaticaly sent
to the robotic telescopes of ASAS. 
We also consider to integrate our system with more sophisticated networks, 
e.g. TALON \cite{TALON}.

Technological progress in image sensors, electronics and computers will undoubtly push
observational astronomy towards huge data streams, which have to be analyzed on-line.
Real time event selection mechanisms, large system issues, maintenance and operation
problems are the areas that need a lot of research and development.
The "{\em $\pi$ of the Sky}" experiment is one of the first steps on this new land.

\section*{Acknowledgments}
We are very much obliged to B.~Paczynski for his support and encouragement.
This project from the very beginning benefits from the experience gained with All Sky
Automatic Survey (ASAS), generously shared with us by its leader, G.~Pojmanski.

\end{document}